\documentstyle[12pt]{article}
\def\iint{\int\!\int}

\def\mint{\int\!\cdots\!\int}
\begin{document}
  \title{Multiplicity Dependence of Identical Particle 
Correlations in the Quantum Optical Approach}
  \author{ { N. Suzuki and M. Biyajima$^{*}$ }
                                     \\
  {\small ${}^{}$Matsusho Gakuen Junior College,}
  {\small Matsumoto 390-1295, Japan        } 
                                     \\
  {\small ${}^{*}$Department of physics,
        Shinshu University, Matsumoto 390-8621, Japan } }
   \date{}
\maketitle
\begin{abstract}
   Identical particle correlations at fixed multiplicity are 
considered in the presence of chaotic and coherent fields.  
  The multiplicity distribution, one-particle momentum density and 
two-particle correlation function are obtained based on the 
diagrammatic representation for cumulants in semi-inclusive events. 
   Our formulation is applied to the analysis of the experimental data 
on the multiplicity dependence of correlation functions reported 
by the UA1 and the OPAL Collaborations.
\end{abstract}
\vspace{5mm}
\section{Introduction}

 In high energy hadron-hadron collisions, Bose-Einstein correlations 
of the identical particles are considered as one of the possible 
measures for the space-time domain where identical particles are 
produced.  One of the theoretical approaches to the Bose-Einstein 
correlations is made on the analogy of the quantum 
optics~\cite{glau63}, where two types of sources, chaotic and 
coherent are introduced.  
A diagrammatical method, based on the Glauber-Lachs 
formula~\cite{glau63}, has been proposed~\cite{biya90} to find the 
higher order Bose-Einstein correlation (BEC) functions in the quantum 
optical (QO) approach.
   In Ref.~\cite{suzu97}, the generating functional (GF) for 
momentum densities is derived in the QO approach, and a diagrammatic 
representation for cumulants is proposed. 

  Up to the present, identical particle correlations in fixed 
multiplicity events are investigated in the case of purely chaotic 
field. Two-particle correlations are analyzed in Ref.~\cite{zajc87} 
by using Monte Carlo methods.  Multiplicity dependence of one-particle 
distributions is discussed in Ref.~\cite{prat93}, and that of two or 
three-particle correlations are considered in Ref.~\cite{chao95}. 

In Ref.~\cite{suzu98}, an outline of our formulation on the particle 
correlations at fixed multiplicity in the QO approach has been 
briefly reported.  General features of multiplicity distributions, 
one-particle distributions and two-particle correlations at 
fixed multiplicity have been also shown. 
   In the present paper, identical particle correlations at fixed 
mutiplicity in the QO approach are considered in detail.  
The diagrammatic representation for cumulants is used to obtain the 
formulas in semi-inclusive events on the analogous way to that in 
inclusive events~\cite{suzu97}.  Furthermore, our formulas are 
applied to the analyses of the experimental data in $p\bar{p}$ 
collisions by the UA1 Collaboration~\cite{busch98}, and 
in $e^+e^-$ collisions by the OPAL Collaboration~\cite{opal96}.

At first, we consider the case when there are no correlations among 
produced particles in the final states.  Then particles in the final 
states are given by a coherent state,
 \begin{equation}
    |\phi\rangle = \exp[ -\frac{1}{2}\int |f(p)|^2 \frac{d^3p}{E}
                 + \int f(p)\,a^{\dag}(p) \frac{d^3p}{E}]|0\rangle.
            \label{eq:gf1}
 \end{equation}
The $n$-particle momentum density in semi-inclusive events is 
defined by
 \begin{eqnarray*}
   \rho_n(p_1,\cdots,p_n)=\frac{1}{\sigma_{\rm inel}}
    E_1\cdots E_n \frac{d^{3n}\sigma_{\rm inel}}{d^3p_1\cdots d^3p_n}
         =\left| \langle 0|a(p_1)\cdots a(p_n)|\phi\rangle\right|^2,
 \end{eqnarray*}
which is reduced to
 \begin{eqnarray}
     \rho_n(p_1,\cdots,p_n)=|f(p_1)|^2\cdots|f(p_n)|^2
              \exp[ -\int |f(p)|^2 \frac{d^3p}{E}].
            \label{eq:gf2}              
 \end{eqnarray}

In the QO approach, the function $f(p)$ is devided into two parts;
  \begin{eqnarray}
       f(p) &=& \sum_{i=1}^M a_i\phi_i(p) + f_c(p),   \label{eq:gf3}
  \end{eqnarray}
where $\phi_i(p)$ and $f_c(p)$ are amplitudes of the $i$th chaotic 
source and a coherent source, and $a_i$ is a random complex number 
attached to the $i$th chaotic source.  In addition, $M$ is the number 
of independent chaotic sources~\cite{suzu97}, which is regarded to be 
infinite in the present paper. The $n$-particle momentum density in 
the QO approach~\cite{biya78} is defined by, 
 \begin{eqnarray}
     \rho_n(p_1,\cdots,p_n)=\left\langle |f(p_1)|^2\cdots|f(p_n)|^2
              \exp[ -\int |f(p)|^2 \frac{d^3p}{E}] \right\rangle_a.
                     \label{eq:gf4}
 \end{eqnarray}
In Eq.~(\ref{eq:gf4}), parenthesis $\langle F\rangle_a$ denotes an 
average of $F$ over the random number $a_i$ with a Gaussian 
weight~\cite{glau63};
 \begin{eqnarray}
    \langle F\rangle_a
      = \prod_{i=1}^M \left( {1 \over {\pi \lambda_i}} \int 
       \exp[-{\left| a_i \right|^2 \over \lambda_i}] 
       d^2 a_i \right) F.  
                      \label{eq:gf5}
 \end{eqnarray}

It should be noticed that the classical (pion) fields are randomized 
in our approach.  On the other hand, each mode of the light is 
randomized in the quantum optics~\cite{glau63}.  After the average 
is taken over the random number $a_i$ in Eq.~(\ref{eq:gf5}), terms 
composed of $a_i^la_i^{*\,m}$ in the function $F$ vanish if $l \neq m$.

The generating functional (GF) for momentum densities in  
semi-inclusive events is defined 
by the following equation,
 \begin{equation}
      Z_{\rm sm}[h(p)] = \sum_{n=1}^\infty {1 \over {n!}} 
        \mint \rho_n (p_1,\cdots,p_n) 
      h(p_1)\cdots h(p_n) {d^3p_1 \over E_1}\cdots {d^3p_n \over E_n}.
                     \label{eq:gf6}
 \end{equation}
From Eqs.~(\ref{eq:gf4}) and (\ref{eq:gf6}), the GF is written 
formally as,
 \begin{equation}
        Z_{\rm sm}[h(p)] = \left\langle \exp \left[
            \int\mid f(p)\mid^2 (h(p)-1) {d^3p \over E} \right] 
              \right\rangle_a.
                  \label{eq:gf7}
 \end{equation}
On the right hand side of Eq.(\ref{eq:gf6}), an additional 
constant $Z_{\rm sm}[h(p)=0]$, which does not affect to the 
momentum densities, is added. 
Inversely, the $n$-particle momentum density in the semi-inclusive 
events is given from the GF as
 \begin{eqnarray*}
   \rho_n(p_1,\cdots,p_n)=\left. E_1\cdots E_n
         \frac{\delta^nZ_{\rm sm}[h(p)]}
        {\delta h(p_1)\cdots \delta h(p_n)}
         \right|_{h(p)=0}.
 \end{eqnarray*}

From the sum rule between semi-inclusive and inclusive 
cross-sections~\cite{koba72}, the GF $Z[h(p)]$ for inclusive events 
is connected to that for semi-inclusive events;
 \begin{equation}
        Z[h(p)] =Z_{\rm sm}[h(p)+1] = \left\langle \exp \left[
            \int\mid f(p)\mid^2 h(p) {d^3p \over E} \right] 
              \right\rangle_a.
                  \label{eq:gf8}
 \end{equation}
The $n$-particle inclusive momentum density is given by
 \begin{eqnarray*}
   \rho_{\rm in}(p_1,\cdots ,p_n)=E_1\cdots E_n \left.
          \frac{\delta^nZ[h(p)]}{\delta h(p_1)\cdots \delta(p_n)}
             \right|_{h(p)=0}.
 \end{eqnarray*}

The explicit form of the GF, Eq.~(\ref{eq:gf8}), is shown in 
Ref.~\cite{suzu97}, and  the higher order BEC 
functions in inclusive events are obtained from it.   
In the following section, we would like to show that we can obtain 
higher order momentum densities in semi-inclusive events 
by analogy with a derivation in inclusive events.

\vspace{5mm} 
\section{Generating functional and cumulant}

In the followings, we slightly change the definition of the GF 
from Eq.(\ref{eq:gf7}) to
 \begin{equation}
        Z_{\rm sm}[h(p)] = c_0\left\langle \exp \left[
            \int\mid f(p)\mid^2 h(p) {d^3p \over E} \right] 
              \right\rangle_a,
                \label{eq:cd1}
 \end{equation}
where the exponential damping factor in 
Eq.~(\ref{eq:gf7}) is replaced by a normalization constant $c_0$. 
 Then, the $n$-particle momentum density in the QO 
approach is given by
 \begin{eqnarray}
   \rho_n(p_1,\cdots,p_2) &=& E_1\cdots E_n \left.
          \frac{\delta^nZ[h(p)]}{\delta h(p_1)\cdots \delta(p_n)}
             \right|_{h(p)=0}  \nonumber \\
             &=&  c_0\,\langle \mid f(p_1)\cdots f(p_n)\mid^2 \rangle_a.
           \label{eq:cd2}
 \end{eqnarray}

The GF $G_{\rm sm}[h(p)]$ for cumulants is defined by the 
following equation,
 \begin{equation}
     G_{\rm sm}[h(p)] \equiv \ln Z_{\rm sm}[h(p)], \label{eq:cd3}
 \end{equation}
and the $n$th order cumulant is given by 
 \begin{equation}
      g_n(p_1,\cdots,p_n) = \left. E_1\cdots E_n
       \frac{\delta^n G_{\rm sm}[h(p)]}
        {\delta h(p_1)\cdots \delta h(p_n)}\right|_{h(p)=0}. 
                               \label{eq:cd4}
 \end{equation}
From Eqs.~(\ref{eq:cd2}), (\ref{eq:cd3}) and (\ref{eq:cd4}), we have 
iteration relations for momentum densities, 
 \begin{eqnarray}
    \rho_1(p_1)&=& c_0g_1(p_1),    \nonumber \\
    \rho_n(p_1,\cdots,p_n) 
       &=& g_1(p_1)\rho_{n-1}(p_2,\cdots,p_n)  \nonumber \\
       &+& \sum_{i=1}^{n-2} \sum 
               g_{i+1}(p_1,p_{j_1},\cdots,p_{j_i}) 
        \rho_{n-i-1}(p_{j_{i+1}},\cdots,p_{j_{n-1}}) \nonumber \\
       &+& c_0g_n(p_1,\cdots,p_n). 
                                              \label{eq:cd5}
 \end{eqnarray}
The second summation on the right hand side of Eq.~(\ref{eq:cd5}) 
indicates that all possible combinations of $(j_1,\cdots , j_i)$ and 
$(j_{i+1},\cdots ,j_{n-1})$ are taken from $(2,3,\cdots, n)$.  
Equation (\ref{eq:cd5}) shows that the $n$-particle momentum density 
$\rho_n(p_1,\cdots,p_n)$ $(n=1,2,\cdots)$ can be evaluated if the 
cumulant $g_n(p_1,\cdots ,p_i)$ $(i=1,2,\cdots ,n)$ is obtained.

The semi-inclusive one-particle and two-particle cumulants are given 
from Eq.~(\ref{eq:cd4}) respectively as
 \begin{eqnarray*}
      g_1(p_1)&=& \langle f(p_1)\rangle_a 
                 =r(p_1,p_1)+c(p_1,p_1),  \\
      g_2(p_1,p_2)&=& \langle f(p_1)f(p_2)\rangle_a  
                      - \langle f(p_1)\rangle_a 
                        \langle f(p_2)\rangle_a \nonumber \\
               &=& |r(p_1,p_2)|^2 + 2{\rm Re}[r(p_1,p_2)c(p_2,p_1)],
 \end{eqnarray*}
where $r(p_1,p_2)$ is a correlation caused by the chaotic sources and 
$c(p_1,p_2)$ is a correlation by the coherent source.  
Those are given by  
 \begin{eqnarray}
    r(p_1,p_2) &=& \sum_{i=1}^M \lambda_i \phi_i(p_1) \phi^*_i(p_2 ), 
                                         \nonumber \\
      c(p_1,p_2) &=&  f_c(p_1) f^*_c(p_2) . 
                  \label{eq:cd6}   
 \end{eqnarray}

As the GF, Eq.(\ref{eq:cd1}), in semi-inclusive events is the same 
with the GF, Eq.(\ref{eq:gf8}), in inclusive events except for the 
normalization factor $c_0$, the cumulants of semi-inclusive events are 
also calculated from the same diagrammatic presentation as those of 
inclusive events~\cite{suzu97}.  
Diagrammatic representation for them up to the fourth order is shown 
in Fig.1.

The $n$th order cumulant is made up of connected terms of the 
correlations $r_{ij}$ of chaotic fields, and those $c_{ij}$ of the 
coherent field.   If $n \ge 3$, the $n$th order cumulant 
$g_n(p_1,\cdots,p_n)$ is simply expressed by using the $n$-gon. 
The $n$th order cumulant consists of two types of terms; one is made 
only of the correlations of the chaotic fields, and the other 
contains the correlations both of the chaotic fields and of the 
coherent field.  One of the terms belonging to the former type is 
given by $r_{12}r_{23}\cdots r_{(n-1)n}r_{n1}$. It can be expressed 
by the circular permutation $(12\cdots n)$ started from 1.
  Other terms can be made from $(12\cdots n)$  by any permutation.  
Therefore, there are $(n-1)!$ different terms in the former type. 
If any one of the correlations $r_{ij}$ of the chaotic fields, 
belonging to the former type, is replaced by the correlation $c_{ij}$ 
of the coherent field,  the terms of the latter type are made.  
Therefore, there are $n!$ different terms in the latter type.   
It should be noted that the half of the terms in the $n$th order 
cumulant are complex conjugates to the other half.

\section{Basic formulas at fixed multiplicity}

In order to calculate momentum densities at fixed multiplicity, 
the $k$-particle momentum density and $k$th order cumulant at 
$n$-particle events ($k\le n$) are defined by the following 
equations, 
respectively
 \begin{eqnarray}
   \rho_n^{(k)}(p_1,\cdots ,p_k)=\frac{1}{(n-k)!}
          \mint \rho_n(p_1,\cdots ,p_k,p_{k+1},\cdots ,p_n)
            \frac{d^3p_{k+1}}{E_{k+1}}\cdots\frac{d^3p_n}{E_n},
                             \nonumber \\
    g_n^{(k)}(p_1,\cdots ,p_k)=\frac{1}{(n-k)!}
          \mint g_n(p_1,\cdots ,p_k,p_{k+1},\cdots ,p_n)
            \frac{d^3p_{k+1}}{E_{k+1}}\cdots\frac{d^3p_n}{E_n}.
                  \label{eq:bf1}
 \end{eqnarray}
The GF for multiplicity distribution $P(n)$ is given from 
Eq.~(\ref{eq:cd1}) if function $h(p)$ is independent from 
momentum $p$;
 \begin{equation}
        Z_{\rm sm}(h) = c_0 \left\langle \exp \left[
            \int\mid f(p)\mid^2 {d^3p \over E}h \right] 
              \right\rangle_a.
                 \label{eq:bf2}
 \end{equation}
The multiplicity distribution is given from Eq.(\ref{eq:bf2});
 \begin{eqnarray}
     P(n)=\frac{1}{n!}\left. \frac{\partial^n Z_{\rm sm}(h)}
            {\partial h^n} \right|_{h=0}
           =\frac{1}{n!}\rho_n^{(0)}.  \label{eq:bf2b}
 \end{eqnarray}
The normalization of the $k$-particle momentum density 
at $n$-particle events is given by
 \begin{eqnarray*}
          \mint \rho_n^{(k)}(p_1,\cdots ,p_k)
            \frac{d^3p_1}{E_1}\cdots\frac{d^3p_k}{E_k}
               =\frac{n!}{(n-k)!}\,P(n).
 \end{eqnarray*}
Then the normalized $k$-particle momentum density 
in $n$-particle events is defined as
 \begin{eqnarray}
    \tilde{\rho}_n^{(k)}(p_1,\cdots ,p_k)&=& \frac{(n-k)!}{n!}
          \frac{\rho_n^{(k)}(p_1,\cdots , p_k)}{P(n)}.
               \label{eq:bf3}
 \end{eqnarray}

In general, the inclusive $k$-particle momentum density is given from 
the semi-inclusive momentum densities as
 \begin{eqnarray*}
           \rho_{\rm in}^{(k)}(p_1,\cdots ,p_k)= \sum_{n=k}
           \rho_n^{(k)}(p_1,\cdots ,p_k),
 \end{eqnarray*}
which satisfies
 \begin{eqnarray*}
          \mint \rho_{\rm in}^{(k)}(p_1,\cdots ,p_k)
            \frac{d^3p_1}{E_1}\cdots\frac{d^3p_k}{E_k}
               =\sum_{n=k} \frac{n!}{(n-k)!}\,P(n).
 \end{eqnarray*}
It should be noted that if the $k$th order inclusive momentum 
densities is integrated over all of the momenta, it becomes 
the $k$th order factorial moment.

From Eqs.~(\ref{eq:cd5}), (\ref{eq:bf1}) and (\ref{eq:bf2b}), the 
multiplicity distribution, and particle densities in semi-inclusive 
events up to the third order can be expressed by the following 
recurrence equations, 
 \begin{eqnarray}
     P(n) &=&\frac{1}{n}
              \sum_{j=1}^n jg_j^{(0)}P(n-j),  
               \label{eq:bf4a} \\
     \rho_n^{(1)}(p_1)&=&
          \sum_{j=1}^n g_j^{(1)}(p_1)P(n-j),  
               \label{eq:bf4b} \\
     \rho_n^{(2)}(p_1,p_2)&=&
                \sum_{j=1}^{n-1} g_j^{(1)}(p_1)\rho_{n-j}^{(1)}(p_2)
          +\sum_{j=2}^{n} g_j^{(2)}(p_1,p_2)P(n-j),
               \label{eq:bf4c} \\
     \rho_n^{(3)}(p_1,p_2,p_3)&=& 
         \sum_{j=1}^{n-2} g_j^{(1)}(p_1)\rho_{n-j}^{(2)}(p_2,p_3)
                       \nonumber \\
         &+&\sum_{j=2}^{n-1} 
           \Bigl\{ g_j^{(2)}(p_1,p_2)\rho_{n-j}^{(1)}(p_3)+
                 g_j^{(2)}(p_1,p_3)\rho_{n-j}^{(1)}(p_2) \Bigr\}
                       \nonumber \\
     &+& \sum_{j=3}^{n} g_j^{(3)}(p_1,p_2,p_3)P(n-j),
               \label{eq:bf4d}
 \end{eqnarray}
where $ P(0)=c_0 $.
On the other hand, cumulants at fixed multiplicity are obtained 
from Fig.1 as
 \begin{eqnarray}
    g_1^{(0)}&=& \Delta_1^{(R)} + \Delta_0^{(S)}, \nonumber \\
   g_n^{(0)}&=&\frac{1}{n}\Bigl[
         \Delta_n^{(R)} +2 \Delta_{n-1}^{(S)}
         + \sum_{j=1}^{n-2}\Delta_{j,n-j-1}^{(T)} \Bigr] ,
            \;\;  n=2,3, \cdots,
               \label{eq:bf5}
 \end{eqnarray}
 \begin{eqnarray}
    g_1^{(1)}(p_1)&=& r(p_1,p_1) + c(p_1,p_1), \nonumber  \\
    g_n^{(1)}(p_1)&=& R_n(p_1,p_1)+2S_{n-1}(p_1,p_1)+
         \sum_{j=1}^{n-2} T_{j,n-j-1}(p_1,p_1), \;\;  n=2,3,\cdots,
               \label{eq:bf6}
 \end{eqnarray}
 \begin{eqnarray}
    g_n^{(2)}(p_1,p_2)&=& \sum_{j=1}^{n-1}R_j(p_1,p_2)R_{n-j}(p_2,p_1) 
            + 2c(p_1,p_2)R_{n-1}(p_2,p_1) 
                           \nonumber   \\
       &+& 2 \sum_{j=1}^{n-2} \Bigl\{ S_j(p_1,p_2)R_{n-j-1}(p_2,p_1)
                +R_{n-j-1}(p_1,p_2)S_j(p_2,p_1) \Bigr\}  
                           \nonumber    \\
      &+& \sum_{j=1}^{n-3}\sum_{l=1}^{n-j-2} \Bigl\{
         T_{j,l}(p_1,p_2)R_{n-j-l-1}(p_2,p_1)
           + R_{n-j-l-1}(p_1,p_2)T_{j,l}(p_2,p_1)\Bigr\}, \nonumber \\
               \label{eq:bf7}
 \end{eqnarray}
 \begin{eqnarray}
    g_n^{(3)}(p_1,p_2,p_3) &=& \sum_{j=1}^{n-2}\sum_{l=1}^{n-j-1}
         \Bigl\{  R_j(p_1,p_2)R_l(p_2,p_3)R_{n-j-l}(p_3,p_1) 
                     \nonumber \\
          &&  + R_j(p_1,p_3)R_l(p_3,p_2)R_{n-j-l}(p_2,p_1) \Bigr\}
                     \nonumber \\
        &+& 2\sum_{j=1}^{n-2}
           \Bigl\{  c(p_1,p_2)R_j(p_2,p_3)R_{n-j-1}(p_3,p_1) 
                     \nonumber \\
           &&  + c(p_1,p_3)R_j(p_3,p_2)R_{n-j-1}(p_2,p_1) 
                     \nonumber \\
          & &  + c(p_2,p_3)R_{n-j-1}(p_3,p_1)R_j(p_1,p_2) \Bigr\}
                     \nonumber \\
       &+& \sum_{j=1}^{n-3}\sum_{l=1}^{n-j-2} \Bigl\{  
           S_j(p_1,p_2)R_l(p_2,p_3)R_{n-j-l-1}(p_3,p_1) 
                     \nonumber  \\
        &&    + S_j(p_1,p_3)R_l(p_3,p_2)R_{n-j-l-1}(p_2,p_1) 
                     \nonumber  \\
        &&  + R_j(p_1,p_2)S_l(p_2,p_3)R_{n-j-l-1}(p_3,p_1) 
                     \nonumber  \\
        &&  + R_j(p_1,p_3)S_l(p_3,p_2)R_{n-j-l-1}(p_2,p_1) 
                     \nonumber  \\
        &&  + R_j(p_1,p_2)R_{n-j-l-1}(p_2,p_3)S_l(p_3,p_1) 
                     \nonumber  \\
        &&  + R_j(p_1,p_3)R_{n-j-l-1}(p_3,p_2)S_l(p_2,p_1) \Bigr\}
                     \nonumber  \\
      &+& \sum_{j=1}^{n-4}\sum_{l=1}^{n-j-3}\sum_{m=1}^{n-j-l-2}
          \Bigl\{  
          T_{j,l}(p_1,p_2)R_m(p_2,p_3)R_{n-j-l-m-1}(p_3,p_1)
                     \nonumber  \\
      &&   +T_{j,l}(p_1,p_3)R_m(p_3,p_2)R_{n-j-l-m-1}(p_2,p_1)
                     \nonumber  \\
       &&+ R_m(p_1,p_2)T_{j,l}(p_2,p_3)R_{n-j-l-m-1}(p_3,p_1)
                     \nonumber  \\
       &&+R_m(p_1,p_3)T_{j,l}(p_3,p_2)R_{n-j-l-m-1}(p_2,p_1)
                     \nonumber  \\
       &&+ R_m(p_1,p_2)R_{n-j-l-m-1}(p_2,p_3)T_{j,l}(p_3,p_1)
                     \nonumber  \\
       && +R_m(p_1,p_3)R_{n-j-l-m-1}(p_3,p_2)T_{j,l}(p_2,p_1)
                                              \Bigr\},
               \label{eq:bf8}
 \end{eqnarray}
where
 \begin{eqnarray}
   R_1(p_1,p_2)&=& r(p_1,p_2),  \nonumber \\
   R_n(p_1,p_2)&=& \int r(p_1,k)R_{n-1}(k,p_2)\frac{d^3k}{\omega},
             \quad n=2,3,\cdots,  \nonumber  \\                 
   S_0(p_1,p_2)&=&c(p_1,p_2),  \nonumber \\
   S_n(p_1,p_2)&=&\int c(p_1,k)R_n(k,p_2)\frac{d^3k}{\omega},
             \quad n=1,2,\cdots,  \nonumber \\
   T_{j,l}(p_1,p_2)&=&\iint R_j(p_1,k_1)c(k_1,k_2)R_l(k_2,p_2)
            \frac{d^3k_1}{\omega_1}\frac{d^3k_2}{ \omega_2},
               \label{eq:bf9}
 \end{eqnarray}
 \begin{eqnarray}
   \Delta_n^{(R)} &=& \int R_n(k,k)\frac{d^3k}{\omega},  \quad
                \nonumber \\
   \Delta_n^{(S)} &=& \int S_n(k,k)\frac{d^3k}{\omega},  \quad
                \nonumber \\
   \Delta_{j,l}^{(T)} &=& \int T_{j,l}(k,k)\frac{d^3k}{\omega}.  
               \label{eq:bf10}
 \end{eqnarray}

In the followings, variables are changed from $(p_{1L},{\bf p}_{1T})$ 
to $(y_1,{\bf p}_{1T})$, with rapidity 
$y_1 = \tanh^{-1}(p_{1L}/E_1)$.
Both correlations $r(p_1,p_2)$ and $c(p_1,p_2)$ are assumed to be real 
and parametrized as,
 \begin{eqnarray}
     r(y_1,{\bf p}_{1T};y_2,{\bf p}_{2T}) &=& p
      \sqrt{\rho(y_1,{\bf p}_{1T})\rho(y_2,{\bf p}_{2T})}
                    \,I(\Delta y, \Delta {\bf p}_{1T}), 
                          \nonumber  \\
     c(y_1,{\bf p}_{1T};y_2,{\bf p}_{2T}) &=& (1-p)
          \sqrt{\rho(y_1,{\bf p}_{1T})\rho(y_2,{\bf p}_{2T})}, 
                           \nonumber \\
    \rho(y_1,{\bf p}_{1T}) &=&  <n_0>\sqrt{\frac{\pi}{\alpha}}
            \frac{\pi}{\beta}
       \exp[ -\alpha\,y_1^2 - \beta\, {\bf p}_{1T}^2 ],  
                           \nonumber  \\
    I(\Delta y, \Delta{\bf p}_T) &=& 
    \exp[ -\gamma_L (\Delta y)^2 - \gamma_T(\Delta {\bf p}_T)^2 ],  
                            \label{eq:bf11}
 \end{eqnarray}
where $p = r(p_i,p_i) / \rho(p_i)$ is called the chaoticity parameter, 
$\Delta y =y_2 -y_1$ and 
$\Delta {\bf p}_T = {\bf p}_{2T} - {\bf p}_{1T}$.
Functions defined by Eqs.~(\ref{eq:bf9}), (\ref{eq:bf10}) and 
(\ref{eq:bf11}) are expressed as
 \begin{eqnarray*}
  R_j(y_1,{\bf p}_{1T},y_2,{\bf p}_{2T})&=& 
       N_j\exp [-A_j(y_1^2+y_2^2)+2C_jy_1y_2]  \nonumber \\
       &\times&  \exp [ - U_j({\bf p}_{1T}^2+{\bf p}_{2T}^2)
                + 2W_j{\bf p}_{1T}\cdot{\bf p}_{2T}],   
                            \label{eq:bf12}
 \end{eqnarray*}
 \begin{eqnarray*}
  S_j(y_1,{\bf p}_{1T},y_2,{\bf p}_{2T})&=& 
       \frac{(1-p)<n_0>\alpha^{1/2}\beta}
       {\sqrt{A_j+\alpha/2}(U_j+\beta/2)}N_j  \nonumber \\
       &\times&  \exp \left[ -\frac{\alpha}{2}y_1^2 - 
          \left(\frac{\alpha}{2} +
           \frac{\alpha \gamma_L}{A_j+\alpha/2} \right) y_2^2 \right],
                          \nonumber   \\
       &\times&  \exp \left[ -\frac{\beta}{2}{\bf p}_{1T}^2 - 
           \left(\frac{\beta}{2} +
         \frac{\beta \gamma_T}{A_j+\beta/2} \right) 
             {\bf p}_{2T}^2 \right],
                                 \label{eq:bf13}
 \end{eqnarray*}
 \begin{eqnarray}
  T_{i,j}(y_1,{\bf p}_{1T},y_2,{\bf p}_{2T}) &=& 
       \frac{(1-p)<n_0>\pi^{3/2}\alpha^{1/2}\beta}
            {\sqrt{(A_i+\alpha/2)(A_j+\alpha/2)}
                   (U_i+\beta/2)(U_j+\beta/2)} N_i N_j  \nonumber \\
    &\times&  \exp \left[ 
          - \left( \frac{\alpha}{2} +
              \frac{\alpha \gamma_L}{A_i+\alpha/2} \right) y_1^2 
          - \left( \frac{\alpha}{2} +
            \frac{\alpha \gamma_L}{A_j+\alpha/2} \right) y_2^2 \right]
                           \nonumber   \\
   &\times&   \exp \left[ 
        - \left( \frac{\beta}{2} + 
          \frac{\beta \gamma_T}{A_i+\beta/2} \right) {\bf p}_{1T}^2 
        - \left( \frac{\beta}{2} +
          \frac{\beta \gamma_T}{A_j+\beta/2} \right) {\bf p}_{2T}^2  
             \right],  \nonumber \\
                            \label{eq:bf14}
 \end{eqnarray}
where  
 \begin{eqnarray}
       A_1 &=& \frac{\alpha}{2} + \gamma_L, \qquad  C_1=\gamma_L, 
                           \nonumber   \\
       A_{j+1} &=& A_1 - \frac{\gamma_L^2}{A_j+A_1}, \quad
       C_{j+1} = \frac{\gamma_LC_j}{A_j+A_1},  
                           \nonumber   \\
       U_1 &=& \frac{\beta}{2}+\gamma_T, \qquad W_1 = \gamma_T, 
                           \nonumber   \\
       U_{j+1} &=& U_1 - \frac{\gamma_T^2}{U_j+U_1}, \quad
       W_{j+1} = \frac{\gamma_TW_j}{U_j+U_1},
                           \nonumber   \\
      N_1 &=& p<n_0> \frac{\alpha^{1/2}\beta}{\pi^{3/2}},  
                           \nonumber   \\
      N_{j+1} &=& \frac{p<n_0>\alpha^{1/2}\beta}
        {\sqrt{A_j+A_1}(U_j+U_1)} N_j. 
                            \label{eq:bf15}
 \end{eqnarray}

\vspace{5mm}
\section{Analyses of experimental data}

Our formulas are applied to the analyses of negatively charged 
particles, or like-sign particles.
Recently, the preliminary data of identical two-particle correlations 
in semi-inclusive events in $p\bar{p}$ collisions at 
$\sqrt{s}=900$ GeV within the pseudo-rapidity interval 
from -3.0 to 3.0 are reported by the UA1 Collaboration~\cite{busch98}. 
As can be seen from Eq.~(\ref{eq:bf4c}), multiplicity distribution 
and one-particle densities are included in the formula of the 
two-particle density, at least the multiplicity distribution of the 
same data sample is required to analyze the two-particle correlation. 
The multiplicity distribution at $\sqrt{s}=900$ GeV is also reported 
by the UA1 group~\cite{ua190}.  However, the data are taken within the 
pseudo-rapidity interval from -2.5 to 2.5.  In the present analysis, 
those data are used to adjust the parameters included 
in our formulation. 

Parameter $\alpha$ is determined from the parametrization of 
one-particle rapidity distribution of Landau's hydrodynamical 
model~\cite{car73}, and $\beta$ is taken from the inclusive 
transverse momentum distribution.  
In the present analysis , we neglect the correlation in the transverse 
momentum space, in other words, $\gamma_T$ is taken to be zero, and 
the calculated value is compared with the data after integrated over 
the transverse momentum.  Therefore, the parametrization of $\beta$ 
does not affect the calculated results. Those values are taken as 
 \begin{eqnarray*}
    \alpha &=&  0.25, \quad     \beta =5.556, \quad \gamma_T=0.
 \end{eqnarray*}

Multiplicity distribution is normalized to satisfy
 \begin{eqnarray}
   \sum_{n=1}^{n_{\rm max}} P(n)=1,   \label{eq:ad1}
 \end{eqnarray}
where $n_{\rm max}=42$ is the maximum multiplicity of the observed 
negatively charged particles.

Other parameters are adjusted to fit the multiplicity 
distribution~\cite{ua190} from $n=1$ to $n=35$ in the following way. 
The chaoticity parameter $p$ is changed from $p=0$ to $p=1.0$ by 
the step 0.1, and other parameters $\langle n_0\rangle$ and 
$\gamma_L$ are determined by the minimum chi-squared method. 
The best fit is given by
 \begin{eqnarray*}
     p=0.8, \quad  \langle n_0\rangle =2.584, \quad   
     \gamma_L =1.394, 
 \end{eqnarray*}
with $\chi^2_{\rm min}/{\rm NDF}=372.4/32$.  
As can be seen from Eq.~(\ref{eq:bf11}), if we keep the relations 
that $\alpha/\gamma_L$=costant and $\beta/\gamma_T$=constant, 
we get the same minimum chi-squared value. The calculated multiplicity 
distribution is compared with the data in Fig.2.  
 For the sake of comparison, we also fit the data by the negative 
binomial distribution, which results in 
$\chi^2_{\rm min}/{\rm NDF}=465.0/33$. 

The normalized one-particle rapidity distribution at $n$-particle 
events is defined by
 \begin{eqnarray}
    \tilde{\rho}_n^{(1)}(y)= 
       \int \tilde{\rho}_n^{(1)} (y,{\bf p}_{T}) d^2{\bf p}_{T},
          \label{eq:ad2}
 \end{eqnarray}
and calculated results for $n=$ 5, 10 and 20 are shown in Fig.3.  
The peak height increases and the width of the distribution becomes 
narrower, as the multiplicity $n$ increases.

In Fig.4, the normalized two-particle rapidity distribution given by
 \begin{eqnarray}
   \tilde{\rho}_n^{(2)}(\Delta y) = \int\int\int 
    \tilde{\rho}_n^{(2)}(y_1,{\bf p}_{1T},y_1 + \Delta y,{\bf p}_{2T})
                dy_1d^2{\bf p}_{1T}d^2{\bf p}_{2T}
          \label{eq:ad3}
 \end{eqnarray}
are shown at $n$=5, 10 and 20. The peak of the distribution also 
becomes higher and it's width becomes narrower, as 
the multiplicity $n$ increases. However the increasing rate is gentler 
than that of the one-particle density.

The normalized two-particle correlation function 
$C_n^{(2-)}(\Delta y)$ at $n$-particle events is defined as
 \begin{eqnarray}
     C_n^{(2-)}(\Delta y) =  \frac{ \int\int\int \tilde{\rho}_n^{(2)} 
            (y_1,{\bf p}_{1T},y_1 + \Delta y,{\bf p}_{2T})
                    dy_1 d^2{\bf p}_{1T} d^2{\bf p}_{2T} }
        { \int\int\int  
          \tilde{\rho}_n^{(1)}(y_1,{\bf p}_{1T}) 
          \tilde{\rho}_n^{(1)}(y_1 + \Delta y, {\bf p}_{2T} ) 
                dy_1d^2{\bf p}_{1T}d^2{\bf p}_{2T} }      - 1.
          \label{eq:ad4}
 \end{eqnarray}
The calculated result on $C_n^{(2-)}(\Delta y)$ at $n$=5, 10 and 20 
are shown on Fig.5.
The multiplicity dependence of $C_n^{(2-)}(\Delta y)$ at 
$\Delta y$=0 and 1.5 are shown in Fig.6, where the preliminary 
experimental data reported by the UA1 Collaboration at $Q=0.1$ GeV 
are also shown 
\footnote{The data on the two-particle correlation are given by 
the variable 4- momentum transfer squared; 
$Q=\sqrt{-(p_1-p_2)^2}$ GeV.  $Q=0$ corresponds to $\Delta y=0$. 
Therefore we compare our calculated results at $\Delta y=0$ with 
the data at the smallest $Q$ value ($Q=0.1$) reported 
by the UA1 Collaboration.}.

In $e^+e^-$ collisions, the OPAL Collaboration published the data on 
multiplicity distributions~\cite{opal92}, and multiplicity 
dependence of two-particle Bose-Einstein correlations~\cite{opal96} 
at 91 GeV.   However,  using the parameters 
adjusted to the observed multiplicity distribution, which is 
close to a Poisson distribution, calculated results on 
$C_n^{(2-)}(\Delta y=0)$ are almost constant and does not show the 
$n$ dependence.   Next, the multiplicity dependence of the  
Bose-Einstein correlations at $Q=0$ GeV, which is estimated from 
the data with $Q \ge 0.05$ GeV, is directly analyzed by our 
formula.  We can fairly well reproduce the $n$ dependence of the data 
with $n_{\rm max}=27$, $\alpha=0.125$, $p=0.55$ and $\gamma_L=10.0$, 
if the minimum values of our calculated results on $C_n^{(2-)}(0)+1$  
are renormalized to 1. The result is shown in Fig.7. 
\vspace{5mm}
\section{Summary and discussions}

   The analytical formulas of multiplicity distribution and particle 
densities in semi-inclusive events are derived from the generating 
functional GF in the presence of the chaotic and coherent fields.  
Formulas are applied to the analysis of the multiplicity dependence 
of two-particle correlations among identical particles in $p\bar{p}$ 
Collisions by the UA1 Collaboration~\cite{busch98} and in $e^+e^-$ 
collisions by the OPAL Collaboration~\cite{opal96}. 
In the formula of two-particle correlation, multiplicity distribution 
and one-particle densities in semi-inclusive events are contained.  
Therefore, to fix the parameters in our formulas, at least the 
observed multiplicity distribution is necessary.  

In $p\bar{p}$ collisions, we adjusted the parameters using the 
multiplicity distribution taken from the different data sample from 
those of the two-particle correlation.  
Our calculated results with the constant chaoticity parameter well 
reproduce the gross features of the multiplicity dependence of 
the data , inspite of the values of $C_n^{(2)}(0)$ are smaller than 
the data at $Q=0.1$ GeV about 20$\%$.  
Some part of the deviation will be deduced to the fact that 
the parameters are determined by fitting the multiplicity 
distribution within a different pseudo-rapidity interval.

In $e^+e^-$ collisions, we analyze the data of the two-particle 
correlation without fitting the multiplicity distribution.  
We can explain the multiplicity dependence of the two-particle 
correlation at $Q=Q_{\rm min}$ GeV observed in the experiment, 
using the constant chaoticity parameter. 

  Calculated results on the normalized two-particle correlation in 
semi-inclusive events show that the peak of the distribution 
becomes lower as the multiplicity increases, 
even if the chaoticity parameter $p$ is constant.  
This behavior is similar to the data of two-particle 
correlations in $p\bar{p}$ collisions by the UA1 Collaboration, 
and in $e^+e^-$ collisions by the OPAL Collaboration.

In this paper, we analyze the data with the same values of chaoticity 
parameter $p$ and the correlation length $\gamma_L$ in rapidity space. 
Our present analyses indicate that the coherent component is not 
negligible, in other words, the values of chaoticity parameters are 
smaller than 1. 
One of the possible candidates for the coherent component is a 
contribution from the decay products of long lived 
resonances~\cite{bowl91}.  Another possibility to reduce the value of 
chaoticity parameter is contanination~\cite{cram96,suzu97}.  
For example, about 20$\%$ of like-sign particles are not pions 
in the OPAL Collaboration~\cite{opal96}. 

When the colliding energy of incident particles increases as in the 
forthcoming RHIC experiment, thousands of identical particles can be 
produced in an event. Then, the production domain of those particles 
can be analysed precisely event by event. 
In general, the values of parameters will change according to the 
multiplicity.  If the fitted values of chaoticity parameter or 
correlation length change suddenly at some multiplicity, 
it will be a possible signature that a threshold of a new phenomenon 
will open at that multiplicity.

\vspace{5mm}

{\bf Acknowledgments}:
M.B. is partially supported by the Japanese Grant-in-Aid for 
Scientific Research from the Ministry of Education, 
Science and Culture (No.09440103) and (No.08304024).  
N.S. thanks for A.Bartl, B.Buschbeck and H.Eggers 
for valuable discussions. 
N.S. also thanks Matsusho Gakuen Junior College for financial support.

%

%
%
%
\begin{description}
 \item[ Figure Captions ]
 \item[ Fig.1] Diagrammatic representation for cumulants in 
   semi-inclusive events. 

  a) Contribution from the chaotic field, $r(p_1,p_2)$, is 
  shown by the solid line with arrow orienting from point 
  $1$ to $2$.  That of the coherent field, $c(p_1,p_2)$, is 
  shown by dotted line with arrow from $1$ to $2$.

b) Diagram for $g_2(p_1,p_2)$.
 
c) Diagram for $g_3(p_1,p_2,p_3)$. All permutations of (2,3) 
   should be taken for $(i,j)$.

d) Diagram for $g_4(p_1,\cdots,p_4)$. Those of (2,3,4) should be 
  taken for $(i,j,k)$

 \item[ Fig.2] Multiplicity distribution observed in $p\bar{p}$ 
  collisions~\cite{opal92} is analyzed by our formula.
  Parameters are determined by the minimum chi-squared method: 
  $p=0.8$, $<n_0>=2.584$ and $\gamma_L=1.394$.

 \item[Fig.3] Normalized one-particle rapidity distributions at 
     fixed multiplicity calculated with $p=0.8$, $<n_0>=2.584$ and 
     $\gamma_L=1.394$.

 \item[Fig.4] Normalized two-particle rapidity distributions at 
     fixed multiplicity calculated with $p=0.8$, $<n_0>=2.584$ 
     and $\gamma_L=1.394$.

 \item[Fig.5] Normalized two-particle correlation functions at 
     fixed multiplicity calculated with 
     $p=0.8$ and $\gamma_L=1.394$.

 \item[Fig.6] Multiplicity dependence of normalized two-particle 
   correlations.   Full circles show the data at $Q$=0.1GeV in 
   $p\bar{p}$  collisions~\cite{busch98}.
   Open circles and open squares are calculated with 
   $p=0.8$ and $\gamma_L=1.394$.
 \item[Fig.7] Multiplicity dependence of normalized two-particle 
   correlations.  Full circles indicate the values at $Q=0$ GeV, 
   estimated from the data with $Q \ge 0.05$ GeV in $e^+e^-$ 
   collisions~\cite{opal96}.  Open circles 
   are obtained from our calculation at $\Delta y=0$.
\end{description}
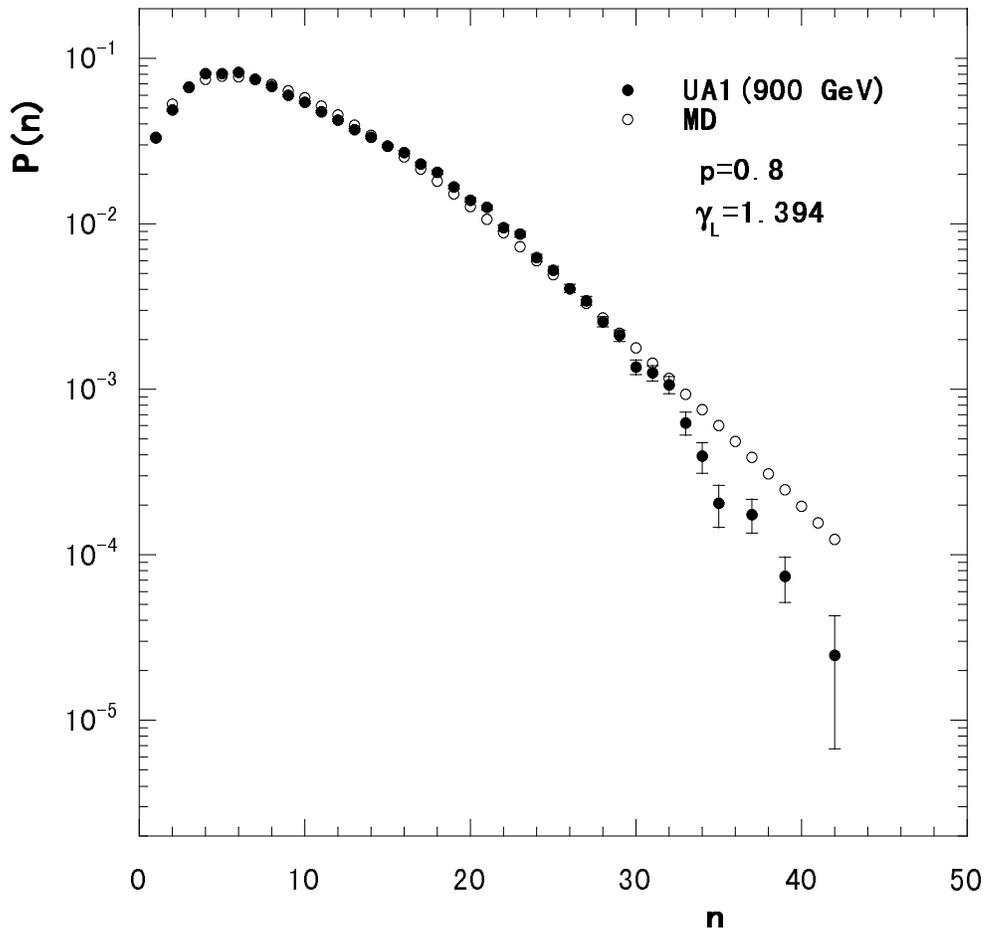
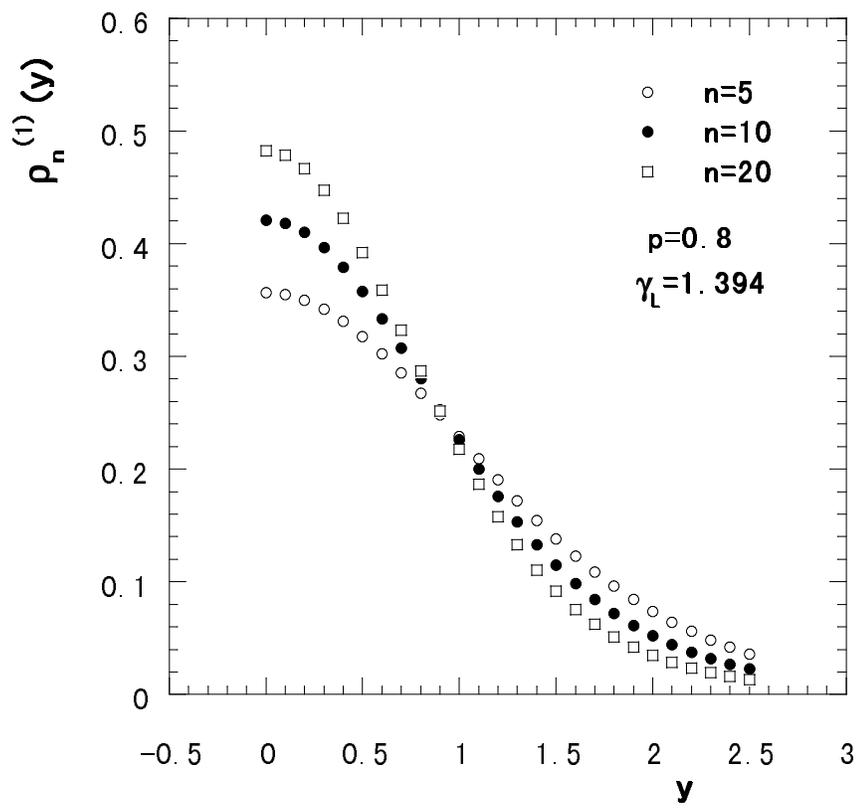
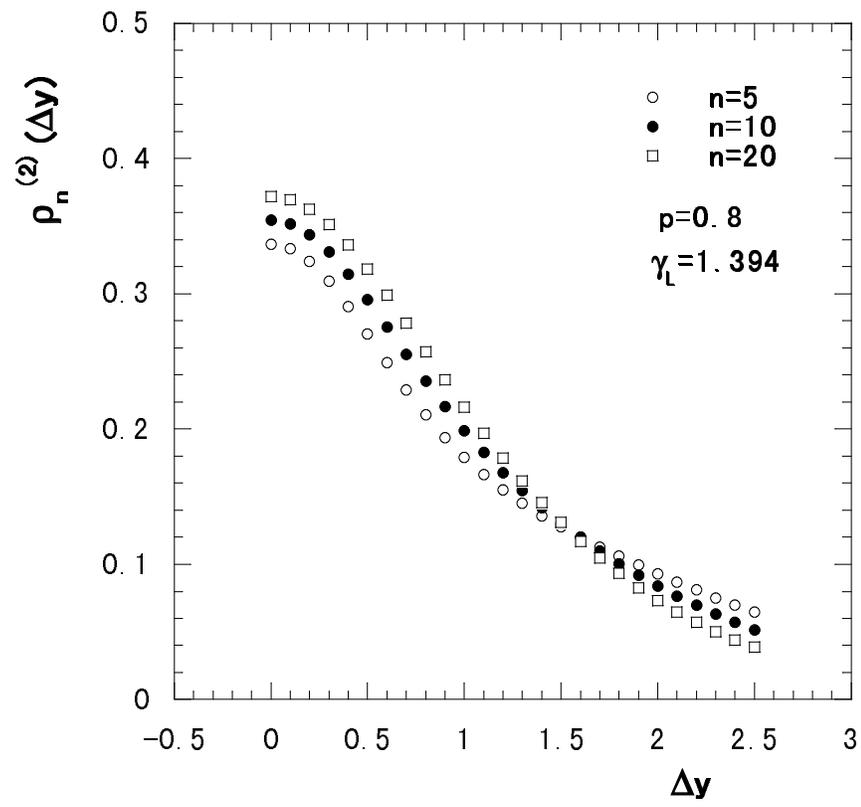
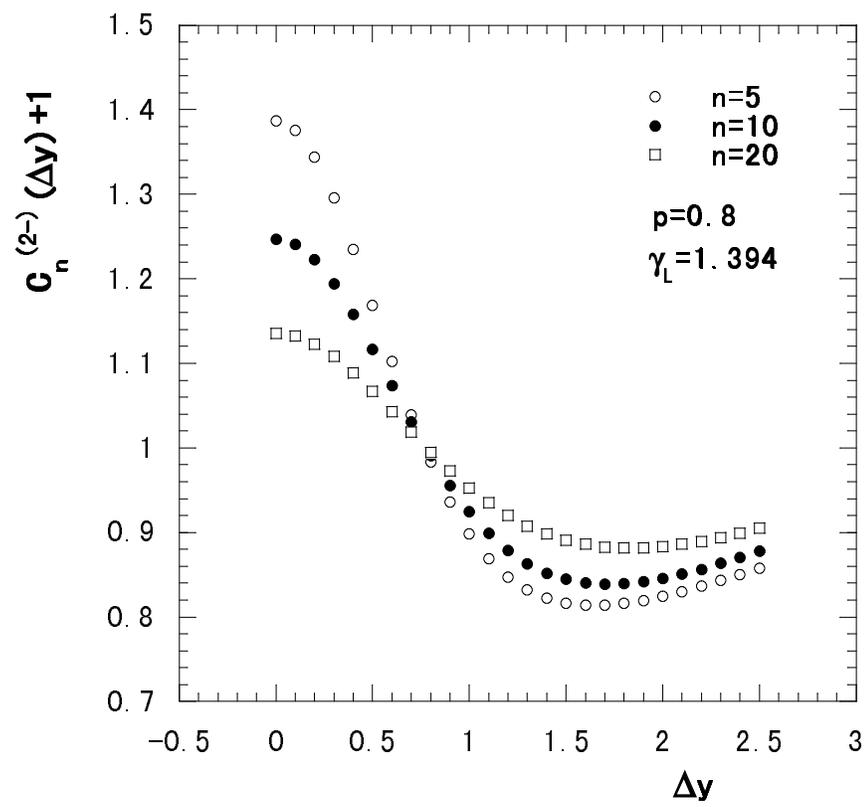
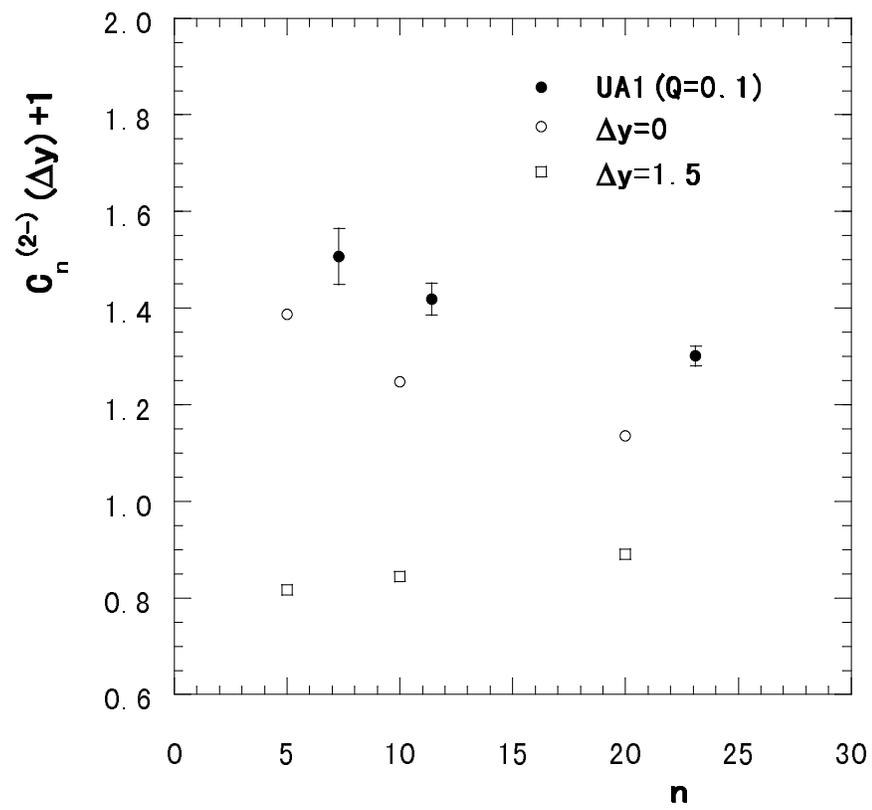
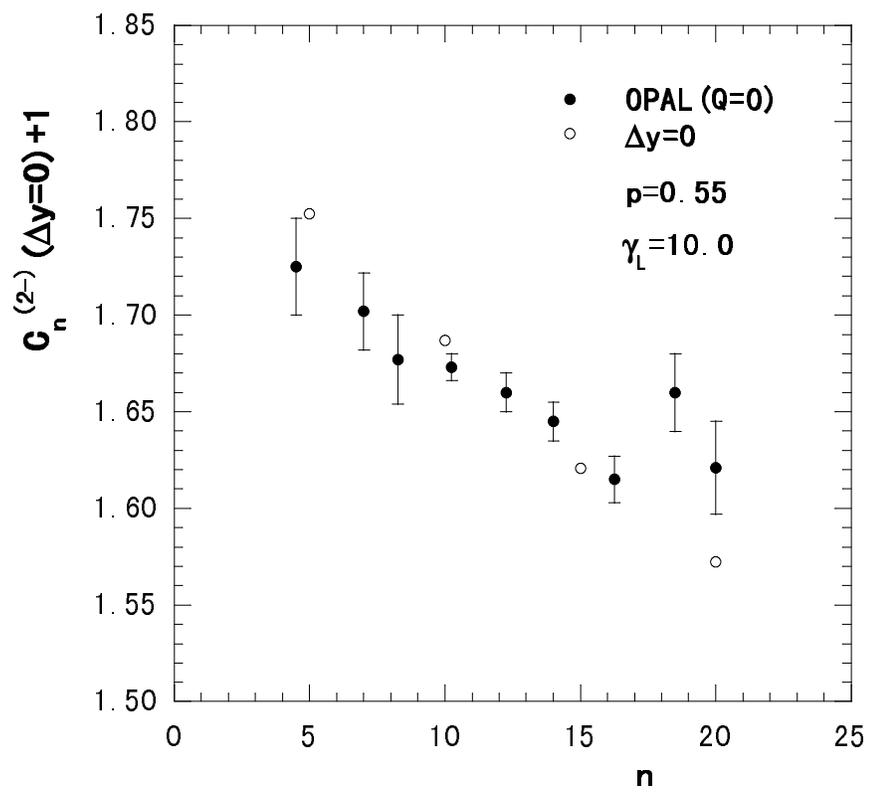
\end{document}